# The topological relationship between the large-scale attributes and local interaction patterns of complex networks

A. Vázquez[1], R. Dobrin[2], D. Sergi[3], J.-P. Eckmann[3,4], Z. N. Oltvai[2] and A.-L. Barabási[1]

[1]Department of Physics, University of Notre Dame, Notre Dame, IN 46556, USA
[2]Department of Pathology, Northwestern University, Chicago, IL 60611, USA
[3]Départment de Physique Théorique and [4]Section de Mathématiques, Université de Genève, Genève, CH-1211, Switzerland

**Abstract**

Recent evidence indicates that the abundance of recurring elementary interaction patterns in complex networks, often called subgraphs or motifs, carry significant information about their function and overall organization. Yet, the underlying reasons for the variable quantity of different subgraph types, their propensity to form clusters, and their relationship with the networks' global organization remain poorly understood. Here we show that a network's large-scale topological organization and its local subgraph structure mutually define and predict each other, as confirmed by direct measurements in five well-studied cellular networks. We also demonstrate the inherent existence of two distinct classes of subgraphs, and show that, in contrast to the low density Type II subgraphs, the highly abundant Type I subgraphs cannot exist in isolation, but must naturally aggregate into subgraph clusters. The identified topological framework may have important implications for our understanding of the origin and function of subgraphs in all complex networks.

**Introduction**

A number of complex biological and non-biological networks were recently found to contain network motifs, representing elementary interaction patterns between small groups of nodes (subgraphs) that occur substantially more often than would be expected in a random network of similar size and connectivity (1, 2). Theoretical and experimental evidence indicates that at least some of these recurring elementary interaction patterns carry significant information about the given network's function and overall organization (1-4). For example, transcriptional regulatory networks of cells (1, 2, 5, 6), neural networks of *C. elegans* (2), and some electronic circuits (2) are all information processing networks that contain a significant number of feed forward loop motifs. However, in transcription-regulatory networks these motifs do not exist in isolation but meld into motif clusters (7), while other networks are devoid of feed-forward loops altogether (2).

In general, all subgraphs have two important properties: their topology and the directionality of their links. In cellular networks, these two properties can be clearly separated from each other. In protein-protein interaction networks all links are by definition non-directional. In contrast, in transcriptional-regulatory networks information flow between a transcription factor and the operon (gene) regulated by it is almost always unidirectional (1, 2). Metabolic networks occupy an intermediate position between these two extremes, as most –but not all- metabolic reactions are reversible under various growth conditions. Despite the difference in the relative role of link directionality, the large-scale organization of the three different network types are quite similar, most being characterized by a scale-free connectivity distribution and hierarchical modularity (8-12). The only exception is the incoming degree distribution (i.e., the number of transcription factors regulating a target gene) of regulatory networks, which decays faster than a power law, as the number of transcription factors that can simultaneously bind to a target gene's promoter region appear to be limited by structural constraints (13).

A coherent understanding of a network's topological and functional organization requires the development of a single framework that can explain the appearance of subgraphs and motifs, the mechanisms responsible for their aggregation into larger superstructures, and their relationship with the universal large-scale features of complex networks. Here we present such a unifying framework by focusing on five well-characterized cellular networks of a prokaryotic and an eukaryotic model organism, the metabolic and transcriptional-regulatory networks of *Saccharomyces cerevisiae* and *Escherichia coli*, respectively, and the protein-protein interaction (PPI) network of *S. cerevisiae*. We show that the subgraph density in these networks can be fully predicted based on knowledge of the two parameters characterizing their global scale-free and hierarchical topology. Furthermore, we demonstrate that a network's large-scale topological organization and its local subgraph structure mutually define and predict each other. We also show the inherent existence of two distinct classes of subgraphs, demonstrating that in contrast to the low density Type II subgraphs, the highly abundant Type I subgraphs cannot exist in isolation, but must naturally aggregate into subgraph clusters. These results imply a fundamental unity in the origin of subgraphs and subgraph clusters in all complex networks.

**Materials and Methods**

**Databases:** The transcriptional regulatory networks of *E. coli* and *S. cerevisiae* (1, 2) are available from http://www.weizmann.ac.il/mcb/UriAlon/. We have studied their undirected representations, where transcription factors and genes are represented by nodes and each regulation-based interaction is replaced by an undirected link. The metabolic networks of *E. coli* and *S. cerevisiae* were obtained from the WIT/ ERGO database (14) (http://igweb.integratedgenomics.com/IGwit/). Metabolites are represented by nodes and undirected links connect each substrate to each product of the same reaction. The protein-protein interaction network of *S. cerevisiae* was obtained from DIP (15) (http://dip.doe-mbi.ucla.edu/). Proteins are represented by nodes and each pair wise protein interaction by an undirected link.

**Results**

**The abundance of subgraphs in cellular networks**

Table 1a lists the density of several *n*-node subgraphs of the five studied intracellular molecular interaction networks: the metabolic and transcriptional-regulatory networks of *Saccharomyces cerevisiae* and *Escherichia coli*, respectively, and the protein-protein interaction (PPI) network of *S. cerevisiae*. Our study is limited to subgraphs with *n* nodes and *m* links, that can be decomposed into a central node with *n*-1 neighbors, the remaining *m-n*+1 links connecting these neighbors to each other. The comparison shown in Table 1a demonstrates that the density of specific subgraphs in the corresponding *E. coli* and *S. cerevisiae* networks are comparable, underscoring the absence of significant differences in the subgraph density between the two organisms. There are notable differences, however, among the different types of molecular interaction webs even within the same organism: the metabolic and the PPI networks display a much higher subgraph density than transcriptional regulatory networks. The observed paucity of certain subgraph types and the abundance of others suggest two possible scenarios for their origin: their number may be largely determined by local functional constraints, such as the desirable signal processing properties of feed-forward motifs (16, 17), or, alternatively, may primarily reflect on the network's topological organization.

To assess the observed paucity of certain subgraph types and the abundance of others we start by focusing on the two key topological parameters of a network's large-scale structure: the degree exponent, $\gamma$ (18), and the hierarchical exponent, $\alpha$ (19). The degree exponent ($\gamma$) characterizes the number of interactions a node is engaged in, capturing the overall inhomogeneity in the connectivity of complex cellular networks: while most molecules are engaged in only a few interactions, a few hubs are linked to a significantly higher number of other molecules (nodes). These wide degree variations are captured by the degree distribution, which for the studied cellular networks follows a power law, $P(k) \sim k^{-\gamma}$ (7, 13, 20-23). In contrast, the hierarchical exponent ($\alpha$) characterizes the networks' innate modularity, indicating that many small, highly interconnected groups of nodes form larger, but less cohesive topological modules (7, 19). This hierarchical modularity is captured by the scaling law (24) $C(k) \sim C_0 k^{-\alpha}$, where $C(k)=2T(k)/k(k-1)$ is the clustering coefficient of a node with *k* links, denoting the probability that a node's neighbors are linked to each other (25), and $T(k)$ is the number of direct links between the node's *k* neighbors. Empirical studies indicate that each cellular network is characterized by a unique pair of ($\gamma,\alpha$) parameters, listed in Table 1b, which were determined from the scaling of $P(k)$ and $C(k)$ functions describing the undirected version of these networks (7, 19).

**Type I and type II subgraphs**

To examine the relationship between these two parameters and the observed subgraph density, we calculated analytically the number $N_{nm}$ of subgraphs with $n$ nodes and $m$ interactions expected for a network of $N$ nodes, in which the nodes -apart from fixed $(\gamma,\alpha)$ parameters-, are randomly connected to each other. As each pair of neighbors of a node with degree $k$ is connected with a probability $C(k) \sim k^{-\alpha}$, the average number of $(n,m)$-subgraphs that pass by a node with degree $k$ scales as $N_{nm}(k) \sim k^{n-1-(m-n+1)\alpha}$. Summing over the degree distribution we obtain the number of $(n,m)$-subgraphs, $N_{nm} \sim N \Sigma_k P(k) N_{nm}(k)$. The convergence of this sum predicts the existence of two subgraph classes. Type I subgraphs are those that satisfy $(m-n+1)\alpha-(n-\gamma)<0$, their number being given by $N^I_{nm} \sim N k_{max}^{-[(m-n+1)\alpha-(n-\gamma)]}$, where $k_{max}$ denotes the degree of the most connected node in the network. Type II subgraphs are those that satisfy $(m-n+1)\alpha-(n-\gamma)>0$, and their number is given by $N^{II}_{nm} \sim N$. As even for finite networks $k_{max} \gg 1$, the typical number of Type I subgraphs is significantly larger than the number of Type II subgraphs ($N^I_{nm}/N^{II}_{nm} \gg 1$). Moreover, for infinite systems ($N \to \infty$) the relative number of Type II subgraphs is vanishingly small compared to Type I subgraphs, as $N^I_{nm}/N^{II}_{nm} \to \infty$. Table 1a supports these predictions, indicating that the density of the subgraphs with a minimal number of connections (extreme Type I) (4,3), (5,4), (6,5), (7,6) is in the range $10-10^5$ ($N^I_{nm} \gg 1$). In contrast, the density of the subgraphs with a maximal number of connections (extreme Type II) (4,6), (5,10), (6,15), (7,21) is either zero or close to zero, and always negligible compared to their Type I counterparts.

The main results of our analysis are summarized in the $(n,m)$ phase diagrams of Fig. 1, in which each square corresponds to a different subgraph. The $(m-n+1)\alpha-(n-\gamma)=0$ condition, predicted to separate the Type I and II subgraphs, appears as stepped yellow phase boundaries in the phase diagrams. For example, for the *E. coli* transcriptional regulatory network with $\alpha=1$ and $\gamma=2.1$ (Table 1b) the phase boundary corresponds to a stepped-line with approximate overall slope $1+1/\alpha=2.0$ and intercept $-1-\gamma/\alpha=-3.1$ (Fig. 1a). The Type II subgraphs are those above this boundary, and should be either absent, or present only in very low numbers in the transcriptional regulatory network. In contrast, the Type I subgraphs below the boundary are predicted to be abundant.

To visually highlight the validity of these predictions we color-coded Fig. 1 according to the normalized count of each subgraph in each cellular network. We find a good agreement between the analytical predictions and the measured subgraph count: the normalized count of the Type I subgraphs below the phase boundary is in the $10^{-2}$-1 range, in contrast with the Type II subgraphs above the predicted boundary, whose normalized count is either zero, or in the $10^{-9}$-$10^{-3}$ range. Comparing Figs. 1a-e indicates that while the stepped phase boundaries for the different cellular networks differ due to the differences in the ($\gamma,\alpha$) exponents (Table 1b), the observed densities in the real networks follow relatively closely the predicted phase boundaries. Occasional local deviations from the predictions can be attributed to the error bars of the ($\gamma,\alpha$) exponents (Table 1b), which allow for some local uncertainties for the phase boundary. Figures 1a-e also indicate that, in agreement with the empirical findings (1-4), each cellular network is characterized by a distinct set of over-represented Type I subgraphs, raising the possibility of classifying networks based on their local structure (4). Yet, the phase diagrams demonstrate that knowledge of two global topological parameters automatically uncovers the local structure of cellular networks, suggesting that a subgraph- or motif-based classification could be equivalent with a classification based on the different ($\gamma,\alpha$) exponents characterizing these networks.

**Subgraphs and motifs**

The concept of *motifs* was recently introduced to denote those subgraphs whose number exceeds by a preset threshold their expected count in a randomized network (1-4). Our results indicate that overrepresented Type I subgraphs are innate topological features of complex networks, and we do not need to invoke a comparison to a randomized graph, nor introduce a threshold parameter to identify them. Indeed, the signature of Type I subgraphs is that their density increases with the number of nodes in the network ($N^I_{nm}/N \to \infty$ as $N \to \infty$), compared with the Type II subgraphs, whose density is independent of the network size ($N^{II}_{nm}/N \to$ const). The existence of the Type II subgraphs is intertwined with the network's global hierarchical topology: the decreasing $C(k)$ reduces the likelihood that the neighbors of a highly connected node are linked to each other, therefore limiting the chance that these nodes participate in highly connected subgraphs. If $C(k)$ were independent of $k$ (i.e., $\alpha$=0), only Type I subgraphs would exist, since in the $\alpha \to 0$ limit the 1+1/$\alpha$ slope of the yellow phase boundary diverges, eliminating all Type II subgraphs. As the absolute count of the subgraphs is the most fundamental quantity for evaluating a local interaction pattern's topological role in a network, we will continue focusing on the direct subgraph count, limiting the discussion on motifs and the role of the randomized reference frame to the Supporting Information. Note that the scaling of the subgraph density with the network size $N$ was already predicted in (26). Yet, the calculation did not take into account the scaling of the clustering coefficient, thus the results are limited to the $\alpha$=0 limit of our predictions. Thanks to the $C(k)$ scaling, however, for realistic $\gamma$ values we predict a new phase, that contain the Type II subgraphs.

**Subgraphs aggregate around hubs**

The very large densities we observe for some Type I subgraphs (Table 1) require us to explain how to distribute as many as $10^{11}$ subgraphs in a network with only $10^3$ nodes. We address this question by calculating the number of distinct subgraphs a given node (gene, metabolite, or protein) participates in. We first focus on the triangle subgraph (3,3), the elementary building block of many higher order subgraphs. A node with $k$ links participates on average in $T(k)=C(k)k(k-1)/2$ triangles. For large $k$ this scales as $T(k) \sim k^{2-\alpha}$. Therefore, the probability that exactly $T$ triangles pass through a node is $P(T) \sim T^{-\delta}$, where $\delta=1+(\gamma-1)/(2-\alpha)$, a power-law dependence that indicates that while the majority of nodes participate in at most one or two triangles, a few nodes take part in a very large number of triangle subgraphs. The monotonic nature of $T(k)$ indicates that the triangles are not distributed uniformly within the network, but tend to aggregate around the hubs. As a node with $k$ links can carry up to approximately $k^2$ triangles, the aggregation around the high $k$ hubs, visible e.g., in Fig. 2a and b, allows the network with a modest number of nodes to absorb a very large number of subgraphs. These calculations can be extended to arbitrary $(n,m)$ subgraphs, in each case predicting a power law for both $T(k)$ and $P(T)$, with exponents that depend on the $(n,m)$ parameters (see the Supporting Information). To test the validity of these analytical predictions we determined numerically $P(T)$ and $T(k)$ for several subgraphs in each of the studied cellular networks. As shown in Figure 2c and d, the results not only support the predicted power law nature of *P*(*T*), but also the numerically determined exponent *δ*, which are in good agreement with the analytically predicted values (Table 1).

The fact that the *P*(*T*) distribution of the individual subgraphs can be uniquely determined by the (*γ*,*α*) exponents has a quite unexpected consequence: it indicates that the relationship between the network's global architecture and its subgraph densities is reciprocal, so that the network's large-scale topology can be uncovered from the inspection of the local subgraph structure. Indeed, by measuring the *P*(*T*) distribution for any *two* subgraphs (e.g., those shown in Figure 2), and using the derived relationship between *δ*, *α* and *γ*, we can determine the *α* and *γ* exponents of the overall network. As the scaling region of *P*(*T*) is more extended than that of *P*(*k*) or *C*(*k*), displaying, for example, over five order of magnitudes of scaling in Fig. 2d, such subgraph-based determination of *γ* and *α* can be at times more precise than the direct fitting of *P*(*k*) and *C*(*k*). Taken together, these findings indicate the *equivalence* of the information obtained from measurements focusing on the local (subgraph based) and global (scale-free and hierarchical) structure of complex networks: a proper characterization of the network's local topology allows us to determine its large scale parameters, or the direct measurement of the network's global statistical features allows us to predict its detailed subgraph structure.

**Subgraph percolation leads to subgraph clusters**

The analytical tools we have developed allow us to uncover how the various subgraphs relate to each other, an issue that is likely to have significant influence on e.g., a particular subgraph's potential functional properties in biological systems. The topological relationship between various subgraphs is illustrated in Figure 3, where we show all nodes participating in several six-node subgraphs (*n*=6) for each of the three studied *S. cerevisiae* cellular networks. The figure indicates that the underrepresented Type II subgraphs, shown on the right, are either absent or form small fragmented islands with only a few nodes. As we move towards the Type I subgraphs shown on the left, we not only observe a rapid increase in the subgraph density, but also a spectacular aggregation process, forcing all the high density Type I subgraphs into a single giant cluster, consisting of thousands to millions of highly interconnected subgraphs.

Our analytical methods permit us to uncover the mechanisms of the observed subgraph aggregation, predicting the existence of a percolation condition given by the equation $(m-n+1)\alpha-(n-2)<0$, such that the subgraphs satisfying this condition should form a giant cluster. The subgraphs that do not satisfy this condition, however, are allowed to break into isolated islands and/or vanish in size. Direct quantitative evidence for the percolation-like transition is provided by the measurement of the relative size of the largest cluster (shown as squares in Fig. 3), indicating that as we move away from the abundant Type I subgraphs, from left to right, the size of the largest cluster shrinks, falling particularly rapidly in the vicinity of the predicted percolation transition. The analytical prediction, shown as a continuous line, indicates a good agreement between the predicted and the measured cluster sizes for the two larger networks (metabolic and protein network). Therefore, these findings indicate that if a node participates in two or more subgraphs, such participation is imposed on the node by the network's topological constraints deriving from the need to distribute a large number of triangles among a finite number of nodes with widely different connectivity.

**Directed subgraphs**

As transcriptional-regulatory interactions and some metabolic reactions are directed, we need to extend our calculations to directed subgraphs as well. For this, we consider directed subgraphs made of $n$ nodes and $m$ directed links that can be decomposed into a central node and $n$-1 in-neighbors ($j$ is an in-neighbor of $i$ if there is a directed link from $j$ to $i$). Among the $m$ directed links, $n$-1 connect the central node to its $n$-1 in-neighbors, while the remaining $m$-$n$+1 directed links connect any two in-neighbors. Whenever there is a link between two in-neighbors they will form, together with the central node, a feed-forward loop (FFL) (1,2). Therefore the problem of finding the number of $(n,m)$ directed subgraphs is equivalent to the undirected case discussed above, after replacing the degree by the in-degree, defined as the number of in-neighbors, the degree distribution by the in-degree distribution $P(k_{in})$, and the clustering coefficient by the FFL clustering coefficient $C_{FFL}$, defined as the number of FFLs passing by a node divided by the maximum number of FFLs that can pass by it. Assuming that $P(k_{in}) \sim k_{in}^{-\gamma}$ and $C_{FFL} \sim k_{in}^{-\alpha}$, our calculations again predict the existence of the Type I and II subgraphs for $(m-n+1)\alpha_{in}-(n-\gamma_{in})<0$ and $(m-n+1)\alpha_{in}-(n-\gamma_{in})>0$, respectively. These results indicate that the distinction between Type I and Type II subgraphs obtained for undirected networks is present in directed networks as well. A complete study of all directed subgraphs can be also completed, but as the discussion of all possible cases is not particular instructive, it is delegated to further work.

**Discussion**

The demonstrated equivalence between the local and global topological organization not only illustrates the importance of taking into account the mathematical realities and constraints when interpreting biological data, but also has a number of important consequences for our understanding of cellular networks. First, it is tempting to conclude that as the large-scale exponents $\alpha$ and $\gamma$ determine the subgraph density, then the global organization has priority over the local one. Such conclusion is a too simplistic, and therefore incorrect. Indeed, a series of studies have indicated that the evolution of the large-scale structure of cellular networks is the consequence of two genome-level mechanisms: gene duplication and the divergence of duplicated molecular interactions due to subsequent mutations (27-32). The combination of these processes allows one to predict the $\alpha$ and $\gamma$ exponents, in agreement with the experimental data (27-32). In contrast, the network's local wiring diagram may be shaped by selection towards subgraphs with desirable functional properties. Therefore, while the global structure reflects the sum of events contributing to the network's growth and buildup, it is often implied that the local properties reflect solely evolutionary selection towards desirable functional traits (1-4). Our results indicate, however, that a sharp distinction between the local and global structure is not justified: determining the large-scale exponents ($\alpha$ and $\gamma$) is equivalent with specifying the number of subgraphs, or providing the distribution of any two subgraphs uniquely identifies the system's large-scale organization and the scaling exponents. Thus, such local processes as gene duplication and subsequent interaction divergence (32) likely determine both the network's large-scale topology ($\alpha$ and $\gamma$) (27-32), and the statistical relevance and density of subgraphs. This common origin of the local and global characteristics is the most likely biological reason for their mathematical equivalence, as neither the density and topology of subgraphs nor the large-scale properties can be dissociated from the evolution of the overall network. Selection for function is likely to play an important role in shaping the directionality and/or strength of the links (e.g., of the molecular interactions for information processing in transcriptional-regulatory networks (1-3)). As our study shows, the inevitable aggregation of Type I subgraphs into clusters is equally important, as it implies that the potential functional properties of statistically abundant subgraphs need also to be evaluated beyond the level of a single subgraph, at the level of subgraphs clusters.

It is important to note that the simplifications we made in the calculations leading to Figs. 1-3 can be relaxed (see the Supporting Information). First, as we have shown above, Type I and II subgraphs can be generalized to directed networks, representing a biologically more relevant approximation for the regulatory and metabolic networks. Second, while Fig. 1 is limited to the subset of $n$-node subgraphs that contain a central node, the results can be generalized to other elementary subgraphs as well, such as those containing cycles of four or more nodes. Subgraphs with a central node are, however, abundant in complex networks with a high clustering coefficient, which is the case for biological networks, and therefore deserve special attention. Finally, the incompleteness of the current maps of cellular networks suggests potentially higher triangle densities than currently detectable. Yet, as long as the missing and false positive interactions are distributed randomly throughout the network, they do not affect our findings. This is supported by the fact that our predictions work equally well for the nearly complete metabolic-, and the incomplete transcriptional-regulatory networks (Figs. 1-3).

In conclusion, the demonstrated mathematical equivalence of a network's large-scale and local, subgraph-based structure underscores the need to understand the properties and evolution of cellular networks as fully integrated systems, where the achievable local changes are inherently intertwined with the network's global organization. Also, the interdependence between the local and global architecture is by no means limited to cellular networks, but is expected to apply to all networked systems, from the World Wide Web to transportation and social networks (8-12, 33). Indeed, preliminary results indicate that the analysis described here can be successfully carried out for the Internet topology and other networks (12, 34, 35), and may have an impact on our understanding of cycles in complex networks, as well (36-37, Vázquez, A. Oliveira, J. G. & Barabási, A.-L. (submitted)). Therefore, while there appears to be significant freedom in the evolution (and subsequent function) of various complex networks, the kind and abundance of local interaction patterns are uniquely characterized by their two global parameters, raising intriguing questions about the role of the local, individual events to shape a network's overall behavior.

**Acknowledgement**

We thank G. Balázsi for comments on the manuscript. Research at the University of Notre Dame and Northwestern University were supported by grants from the U.S. Department of Energy, the NIH , and the NSF. Correspondence and request for materials should be sent to A.-L.B. (E-mail: alb@nd.edu).

**(a)**

| (*n,m*) | | Transcription | | Metabolic | | Protein Interaction |
|---|---|---|---|---|---|---|
| | | *E. coli* | *S. cerevisiae* | *E. coli* | *S. cerevisiae* | *S. cerevisiae* |
| **(3,2)** |  | 12 | 19 | 101 | 72 | 70 |
| **(3,3)** |  | 0.30 | 0.31 | 5.0 | 5.8 | 4.1 |
| **(4,3)** |  | 169 | 220 | 4412 | 2041 | 2395 |
| **(4,6)** |  | 0.00 | 0.00 | 0.44 | 0.77 | 0.97 |
| **(5,4)** |  | 2492 | 2587 | $2.1 \times 10^5$ | $5.9 \times 10^4$ | $1.2 \times 10^5$ |
| **(5,10)** |  | 0.00 | 0.00 | 0.055 | 0.20 | 0.66 |
| **(6,5)** |  | $3.2 \times 10^4$ | $2.8 \times 10^4$ | $8.8 \times 10^6$ | $1.5 \times 10^6$ | $5.7 \times 10^6$ |
| **(6,15)** |  | 0.00 | 0.00 | 0.00 | 0.03 | 0.36 |
| **(7,6)** |  | $3.4 \times 10^5$ | $2.7 \times 10^5$ | $3.5 \times 10^8$ | $3.7 \times 10^7$ | $2.4 \times 10^8$ |
| **(7,21)** |  | 0.00 | 0.00 | 0.00 | 0.00 | 0.00 |

**(b)**

| | | | | | | |
|---|---|---|---|---|---|---|
| γ | | 2.1±0.3 | 2.0±0.2 | 2.0±0.4 | 2.0±0.1 | 2.4±0.4 |
| α | | 1.0±0.2 | 1.0±0.2 | 0.8±0.3 | 0.7±0.3 | 1.3±0.5 |
| β | **meas.** | 1.0±0.2 | 0.8±0.2 | 1.1±0.2 | 1.4±0.2 | 0.7±0.2 |
| | **pred.** | 0.97 | 0.95 | 1.2 | 1.3 | 0.7 |
| δ | **meas.** | 2.1±0.2 | 2.2±0.2 | 1.8±0.2 | 1.7±0.2 | 2.3±0.2 |
| | **pred.** | 2.0 | 1.9 | 1.8 | 1.8 | 3.0 |

**Table 1. Local and global properties of cellular networks:** Panel **a** shows the relative count $N_{nm}/N$ of the least and most connected subgraphs in each of the five studied cellular networks, where $N_{nm}$ represents the number of the given (*n,m*) subgraph found in the network, and $N$ is the total number of nodes in the network. The first and second columns list the subgraph codes and show a representative topology. Panel **b** lists the γ and α exponents for each of the studied cellular networks, determined from a direct fit to the $P(k)$ and $C(k)$ functions of the undirected network representation (see Supporting Information). We also provide the measured and predicted values of the β and δ exponents, characterizing the average number of triangle (3,3) motifs in which a node with $k$ links participates ($T(k) \sim k^\beta$) and the distribution of the number of triangle motifs in which a node participates ($P(T) \sim T^{-\delta}$).

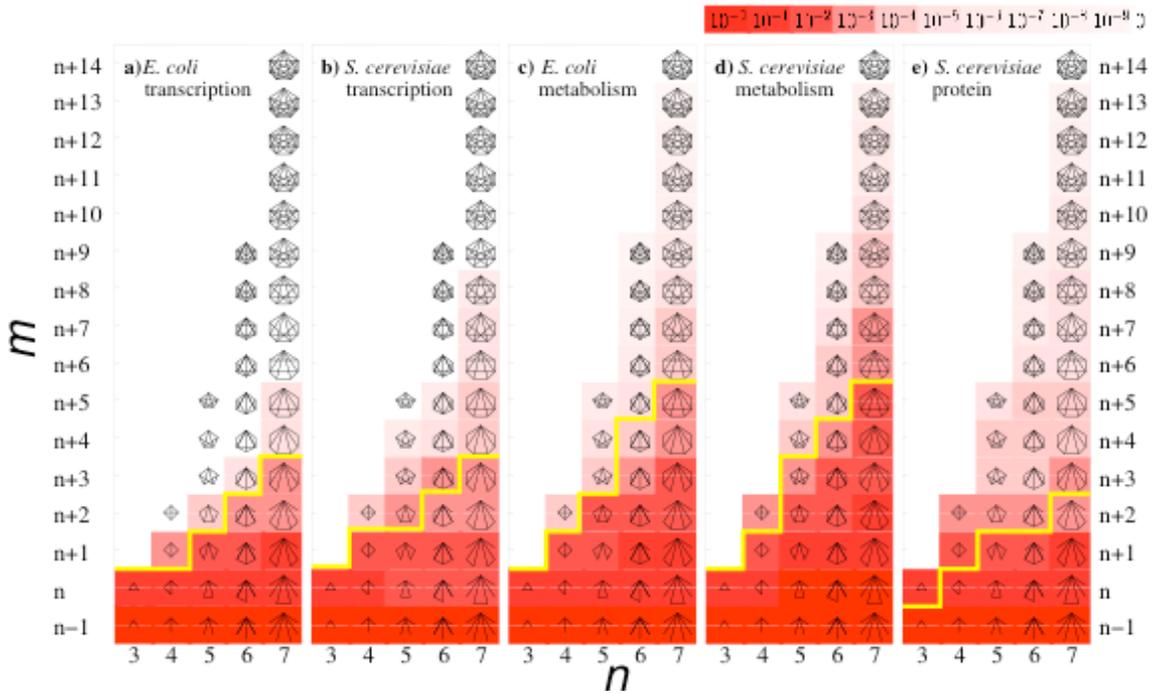

**Figure 1** **Subgraph phase diagrams:** The phase diagrams organize the subgraphs based on the number of nodes ($n$, horizontal axis) and the number of links ($m$, vertical axis), each discrete point explicitly depicting the corresponding subgraph. The stepped yellow line corresponds to the predicted phase boundary separating the abundant Type I subgraphs (below the line) from the constant density Type II subgraphs (above the line). The background color is proportional to the relative subgraph count $C_{nm}=N_{nm}/\Sigma_s N_{ns}$ of each $n$-node subgraph, the color code being shown in the upper right corner. Note that some ($n,m$) points in the phase diagram may correspond to several topologically distinguishable subgraphs. For simplicity, we depict only one representative topology in such cases. As the yellow phase boundary depends on the $\gamma$ and $\alpha$ exponents of the corresponding network, each phase diagram is slightly different. Yet, there is a visible similarity between the networks of the same kind: the phase diagrams of the two transcription or the two metabolic networks are almost indistinguishable.

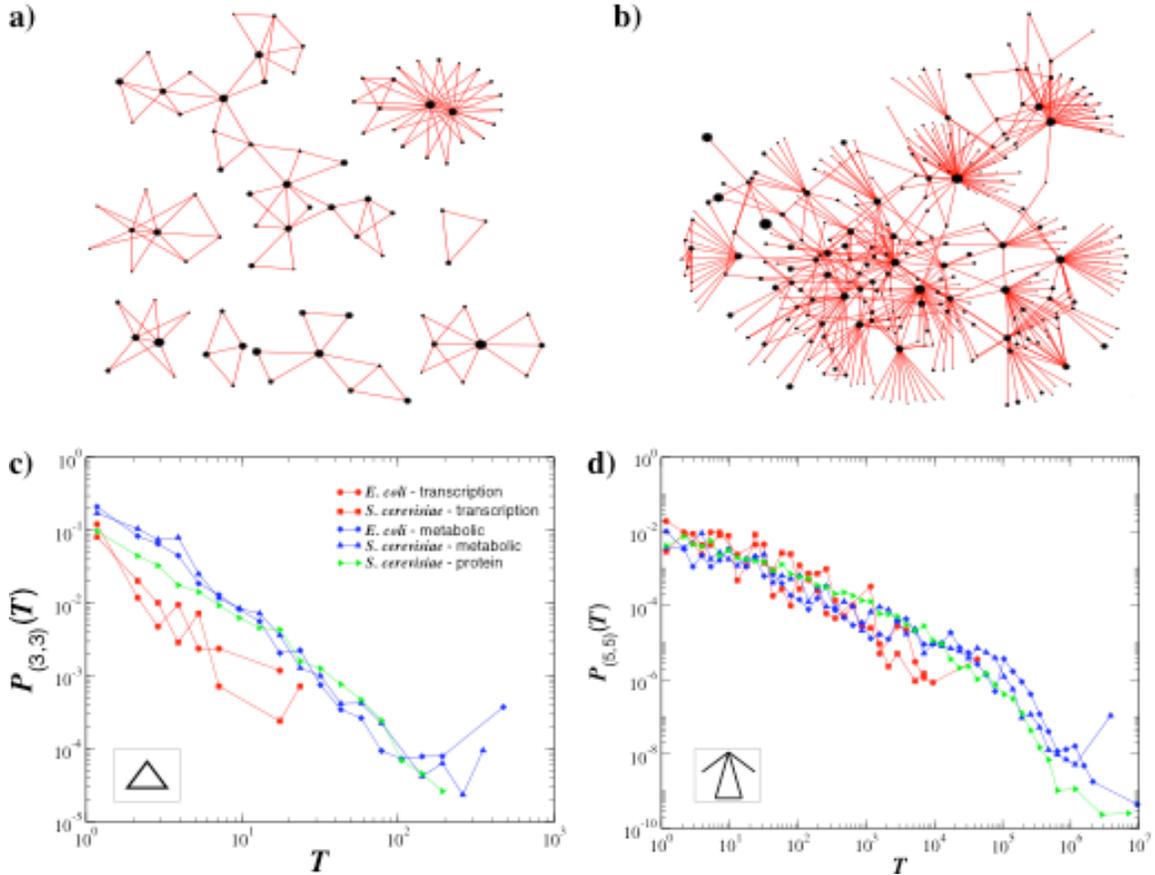

**Figure 2    Subgraph distributions in cellular networks:**

Panels **a** and **b** show all nodes in the *S. cerevisiae* transcription regulatory network that participate in triangle (3,3) and (5,5) subgraphs (depicted in the insets of **c** and **d**). The size (area) of each node is drawn proportional to its degree *k* in the full network, indicating that subgraphs tend to aggregate around the hubs. Indeed, while there are hubs that have only a few subgraphs around them, in most cases subgraph aggregation is seen only around highly connected nodes. Note that the (3,3) subgraph of *S. cerevisiae* is above the percolation boundary (Fig. 1e), and therefore they are broken into small islands. In contrast, the (5,5) subgraph is well below the boundary, forming a fully connected giant component, with no isolated subgraphs, as predicted. The bottom panels show the *P(T)* distribution of the number of **(c)** (3,3) and **(d)** (5,5) subgraphs passing by a node, where the different colors corresponding to the different cellular networks and *T* denotes the number of subgraphs of a selected kind passing by a given node. The plot indicates that for both subgraphs *P(T)* approximates a power law $P(T) \sim T^{-\delta}$. Note the quite extended scaling regimes for some networks: for example for the (5,5) subgraph the scaling extends over four-five orders of magnitude. The $\delta$ exponents measured and predicted for each network are summarized in Table 1b and the Supporting Information.

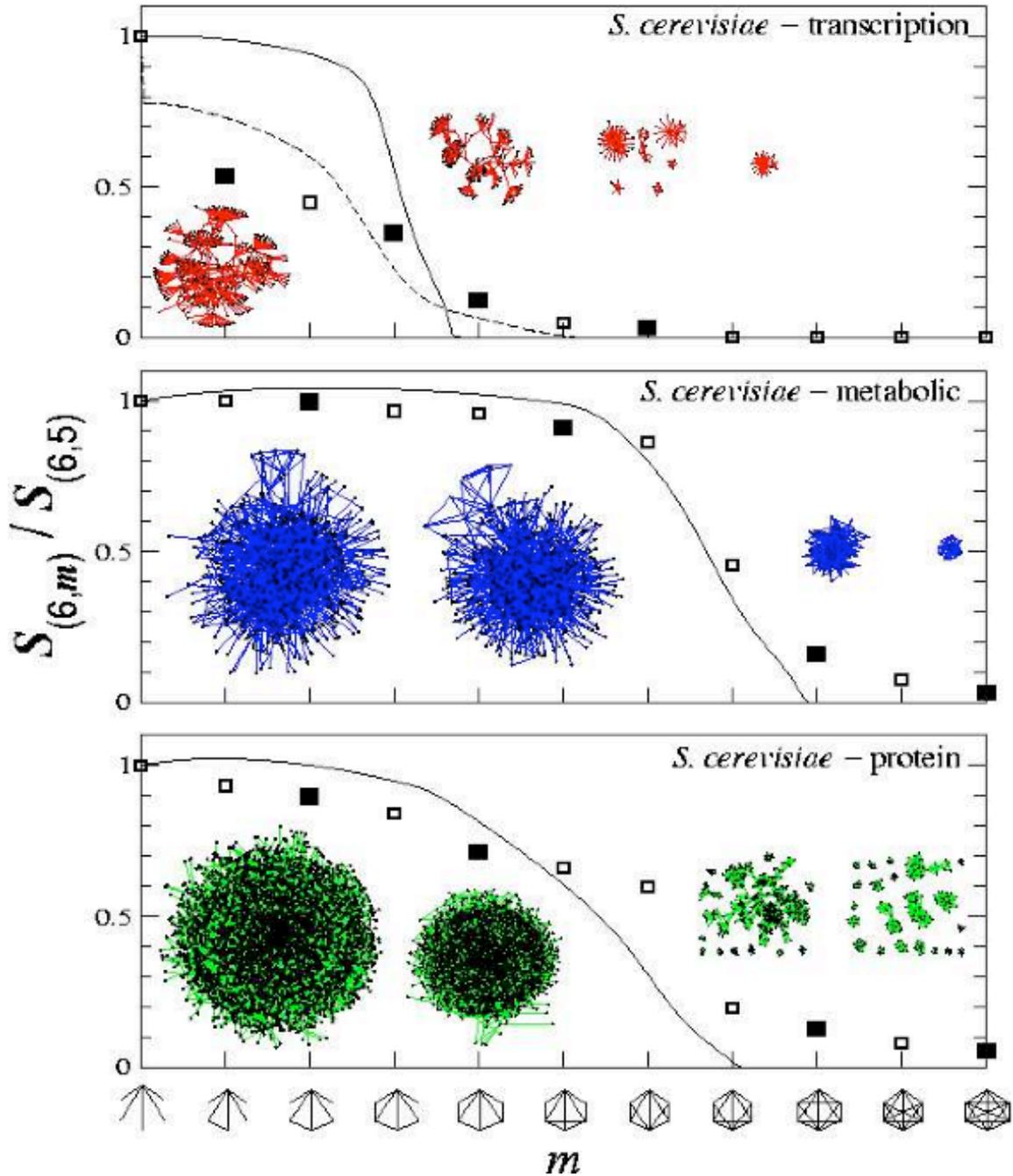

**Figure 3.    Subgraph aggregation and percolation:** The horizontal axis shows the sequence of $n=6$ subgraphs, the number of links ($m$) increasing from left to right. The vertical axis corresponds to the relative size of the largest cluster for the subgraphs shown on the horizontal axis, being determined by the $S_{(6,m)}/S_{(6,5)}$ ratio, where $S_{(6,m)}$ represents the number of nodes participating in ($n=6,m$) subgraphs and $S_{(6,5)}$ represents the total number of nodes participating in the first and most abundant subgraph of the $n=6$ subgraph family. The square symbols represent the measured value of the $S_{(6,m)}/S_{(6,5)}$ ratio for the *S. cerevisiae* networks listed in the upper right corner, indicating that the relative size of the subgraph cluster shrinks from close to one to zero as we move from the highly abundant

Type I subgraphs to the low density Type II subgraphs. The topological consequences of the predicted transition can be seen on the inserted network maps, each corresponding, in order, to the four filled symbols. The sequence of maps demonstrates that while the Type I subgraphs all aggregate into a giant subgraph cluster, as we move towards the Type II subgraphs, the cluster shrinks rapidly in the vicinity of the predicted percolation transition, and disappears by either shrinking to close to zero size (see e.g. the metabolic network) or by breaking into many small islands, which also disappear by further shrinking (see e.g. the transitional-regulatory and protein interaction networks). The continuous line, corresponding to our analytical prediction for the relative cluster size is in quite good agreement with the measured curve for the relatively large protein interaction and metabolic networks. The particular shape of the curve depends, however, on the functional form we use for $C(k)$. For example, the continuous curves were obtained using the analytic approximation $C(k)=C_0/[1+(k/k_0)^\alpha]$. In contrast, the agreement for the transcriptional-regulatory network can be significantly improved by replacing this fit with the directly measured $C(k)$ (dashed line), reproducing even the sharp drop for the relative density of the least connected cluster (first symbol in the top panel).